\documentclass[aps,prb,showpacs,groupedaddress,amssymb,twocolumn]{revtex4}
\usepackage{bm,amsmath}
\usepackage[dvips]{graphicx,color}

\begin{document}

\title{Superconducting phase qubit coupled to a nanomechanical resonator: Beyond the rotating-wave approximation}

\author{Andrew T. Sornborger,$^1$ Andrew N. Cleland,$^2$ and Michael R. Geller$^3$}

\affiliation{$^1$Department of Mathematics, University of Georgia, Athens, Georgia 30602-2451 \\
$^2$Department of Physics, University of California, Santa Barbara, California 93106 \\
$^3$Department of Physics and Astronomy, University of Georgia,  Athens, Georgia 30602-2451}

\date{July 14, 2004}

\begin{abstract}
We consider a simple model of a Josephson junction phase qubit coupled to a solid-state nanoelectromechanical resonator. This and many related qubit-resonator models are analogous to an atom in an electromagnetic cavity. When the systems are weakly coupled and nearly resonant, the dynamics is accurately described by the rotating-wave approximation (RWA) or the Jaynes-Cummings model of quantum optics. However, the desire to develop faster quantum-information-processing protocols necessitates approximate, yet analytic descriptions that are valid for more strongly coupled qubit-resonator systems. Here we present a simple theoretical technique, using a basis of dressed states, to perturbatively account for the leading-order corrections to the RWA. By comparison with exact numerical results, we demonstrate that the method is accurate for moderately strong coupling, and provides a useful theoretical tool for describing fast quantum information processing. The method applies to any quantum two-level system linearly coupled to a harmonic oscillator or single-mode boson field. 
\end{abstract}


\maketitle
\clearpage

\section{INTRODUCTION}

Josephson junctions have been shown to be effective qubit elements for solid-state quantum computing architectures.\cite{Nakamura 1999,Vion etal 2002,Yu 2002,Martinis 2002,Berkley et al 2003,Yamamoto etal 2003} Several proposals for multi-qubit coupling introduce electromagnetic\cite{Shnirman etal 1997,Makhlin etal 1999,Buisson and Hekking 2001,Smirnov and Zagoskin unpublished 2002,Blais etal 2003,Plastina and Falci 2003,Zhu etal 2003,Girvin etal preprint 2003,Blais etal preprint 2004} or mechanical\cite{Cleland and Geller 2004,Geller and Cleland} resonators, or other oscillators,\cite{Marquardt and Bruder 2001,Plastina etal 2001,Hekking etal unpublished 2002} to mediate interactions between the qubits. Such resonator-based coupling schemes have additional functionality resulting from the ability to tune the qubits relative to the resonator frequency, as well as to each other. These qubit-resonator systems are analogous to one or more tunable few-level atoms in an electromagnetic cavity, and the dynamics is often accurately described by the rotating-wave approximation (RWA) or Jaynes-Cummings model of quantum optics.\cite{Scully book}

For a qubit with energy level spacing $\Delta \epsilon$ coupled with strength $g$ to a resonator with angular frequency $\omega_0$ and quality factor $Q$, the RWA is valid when both $|\omega_0 - \Delta \epsilon / \hbar | \ll \omega_0/Q$ and $g \ll \Delta \epsilon$. However, the resonant Rabi frequency, which is proportional to $g$, is then much smaller than the qubit frequency $\Delta \epsilon /\hbar$. Therefore, restricting $g$ to be in the simpler weak coupling regime leads to quantum information processing that is slower than necessary, allowing fewer operations to be performed during the available quantum coherence lifetime.

The threshold theorem \cite{Aharonov and Ben-Or,Gottesman,Kitaev,Preskill} states that if the component failure probability $p$ is below some threshold $p_{\rm th}$, a computation with an error probability bounded by $\eta$ may be accomplished, provided a sufficient number of quantum gates are used for fault-tolerant encoding. In practice, it will be important to have $p$ as small as possible. To approach this limit, we wish to study qubit-resonator systems with stronger coupling (larger $g$) than may be correctly described by the RWA. This will allow us to consider faster switching times for qubit-resonator gates, and to understand to what extent the coupling may be increased while still retaining good fidelity.

In this paper, we use a basis of dressed states\cite{Meystre and Sargent book} to calculate the leading-order corrections to the RWA for a Josephson junction phase qubit coupled to a solid-state nanoelectromechanical resonator, or for any other model of a two-level system linearly coupled to a single-mode boson field. By comparison with exact numerical results, we demonstrate that the method is accurate for moderately strong coupling and provides a useful theoretical tool for describing fast quantum information processing.

\section{JUNCTION-RESONATOR DYNAMICS IN THE DRESSED-STATE BASIS}

\subsection{Qubit-resonator Hamiltonian}

\begin{figure}
\includegraphics[width=8.0cm]{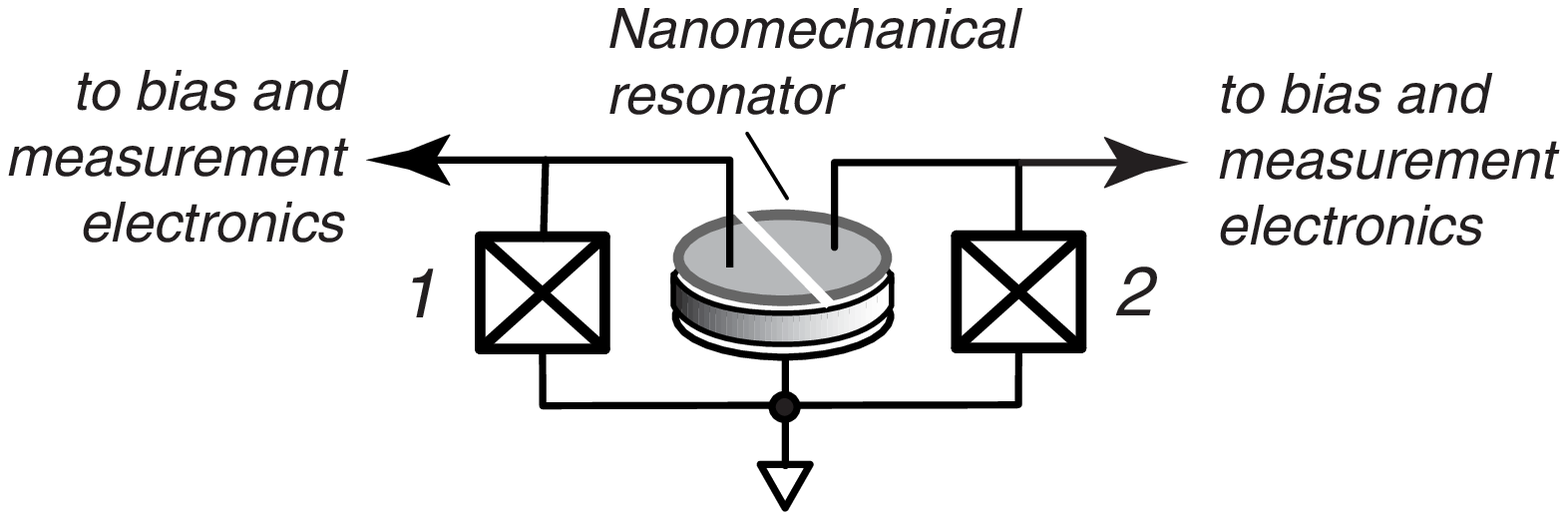}
\caption{\label{architecture figure}Two current-biased Josephson junctions (crossed boxes) coupled to a piezoelectric disc resonator.}
\end{figure}

The Hamiltonian that describes the low-energy dynamics of a single large-area, current-biased Josephson junction, coupled to a piezoelectric nanoelectromechanical disk resonator, can be written as\cite{Cleland and Geller 2004,Geller and Cleland}
\begin{equation}
H = \sum_m \epsilon_m c_m^\dagger c_m + \hbar \omega_0 a^\dagger a - i g \sum_{mm'} x_{mm'} c_m^\dagger c_{m'} (a-a^\dagger),
\label{full Hamiltonian}
\end{equation}
where the $\{c_m^\dagger \}$ and $\{c_m\}$ denote particle creation and annihilation operators for the Josephson junction states $(m \! = \! 0,1,2,\dots), a$ and $a^\dagger$ denote ladder operators for the phonon states of the resonator's dilatational (thickness oscillation) mode of frequency $\omega_0$, $g$ is a coupling constant with dimensions of energy, and $x_{mm'} \equiv \langle m | \delta | m' \rangle$. The value of $g$ depends on material properties and size of the resonator, and can be designed to achieve a wide range of values.\cite{Cleland and Geller 2004,Geller and Cleland} An illustration showing two phase qubits coupled to the same resonator is given in Fig.~\ref{architecture figure}.

For simplicity we will consider only two levels in a single junction; generalization of our method to more than two junction states is cumbersome but straightforward.\cite{generalization footnote} However, all possible phonon-number states are included. The Hamiltonian may then be written as the sum of two terms, $H = H_{\rm JC} + V$. The first term, 
\begin{eqnarray}
H_{\rm JC} &\equiv&  \epsilon_0 \, c_0^\dagger c_0+ \epsilon_1 \, c_1^\dagger c_1 + \hbar \omega_0 \, a^\dagger a \nonumber \\
&-& i g x_{01} [ c_1^\dagger c_0 a - c_0^\dagger c_1 a^\dagger],
\end{eqnarray}
is the exactly solvable Jaynes-Cummings Hamiltonian, the eigenfunctions of which are known as dressed states. We will consider the second term, 
\begin{eqnarray}
V &\equiv&  - i g \big[ x_{00}  c_0^\dagger c_0 (a-a^\dagger) +  x_{01} c_0^\dagger c_1 a  \nonumber \\
&-& x_{01} c_1^\dagger c_0 a^\dagger + x_{11}  c_1^\dagger c_1 (a-a^\dagger)\big],
\end{eqnarray}
as a perturbation. The RWA applied to the Hamiltonian $H$ amounts to neglecting $V$. Therefore, perturbatively including $V$ is equivalent to  perturbatively going beyond the RWA.

\subsection{Dressed states}

It will be useful to define a set of Rabi frequencies according to
\begin{equation}
\Omega_j(\omega_{\rm d}) \equiv \sqrt{ [\Omega_j(0)]^2 + \omega_{\rm d}^2}, \ \ \ \ \ j=0,1,2,\dots
\end{equation}
where
\begin{equation}
\Omega_j(0) \equiv (j+1)^{1 \over 2} \times 2 g |x_{01}|/\hbar
\end{equation}
are the {\it resonant} Rabi frequencies for a qubit coupled to an oscillator containing $j$ phonons, and where 
\begin{equation}
\omega_{\rm d} \equiv \omega_0 - \Delta \epsilon /\hbar 
\end{equation} 
is the resonator-qubit detuning frequency. The vacuum $(j \! = \! 0)$ Rabi frequency on resonance is $\Omega_0(0) = 2 g |x_{01}|/\hbar$.

The eigenstates of $H_{\rm JC}$, or the dressed states, are labeled by the nonnegative integers $j=0,1,2,\dots$ and a sign $\sigma= \pm 1$,
\begin{equation}
\big|\psi^\sigma_j \big\rangle \equiv {[\Omega_j(\omega_{\rm d}) + \sigma \omega_{\rm d}] \big|0,j+1 \big\rangle - i \sigma \Omega_j(0) \big|1,j \big\rangle \over \sqrt{2 \, \Omega_j(\omega_{\rm d}) \, [\Omega_j(\omega_{\rm d}) + \sigma \omega_{\rm d}]}}, 
\label{JC eigenstates}
\end{equation}
where $|mn\rangle \equiv |m\rangle_{\rm JJ} \otimes |n\rangle_{\rm JJ}$ are the eigenstates of the uncoupled system. These states, together with $|00\rangle$, form a complete basis. The energies are
\begin{equation}
W^\sigma_j \equiv \epsilon_0 + (j+1) \hbar \omega_0 -  {\hbar \omega_{\rm d} \over 2} +  \sigma {\hbar \Omega_j(\omega_{\rm d})
\over 2},
\label{JC energies}
\end{equation}
and $H_{\rm JC} |00\rangle = 0$.
On resonance, these reduce to
\begin{equation}
|\psi^\sigma_j\rangle \rightarrow {|0,j+1\rangle - i \sigma | 1,j\rangle \over \sqrt{2}}, \ \ \ \ \ (\omega_{\rm d}=0)
\label{resonant JC eigenstates}
\end{equation}
and
\begin{equation}
W^\sigma_j \rightarrow  \epsilon_0 + (j+1) \hbar \omega_0 + \sigma \sqrt{j+1} \ { \hbar \Omega_0(0) \over 2}. \ \ \ \ \ (\omega_{\rm d}=0)
\label{resonant JC energies}
\end{equation}
Below we will restrict ourselves exclusively to the resonant case.

In what follows, we will need the matrix elements of $V$ in the dressed-state basis, which are given by
\begin{eqnarray}
\langle \psi_{j}^{\sigma} |V| \psi_{j'}^{\sigma'} \rangle &=& -{i g \over 2} \bigg[ \sqrt{j+2} \, x_{00} \, \delta_{j+1,j'} - \sqrt{j+1} \, x_{00} \, \delta_{j,j'+1} \nonumber \\
&-& i \sigma \sqrt{j} \, x_{01} \, \delta_{j,j'+2} -  i \sigma' \sqrt{j+2} \, x_{01} \, \delta_{j+2,j'} \nonumber \\
&+& \sigma \sigma' \, x_{11} \bigg( \sqrt{j+1} \, \delta_{j+1,j'} - \sqrt{j} \, \delta_{j,j'+1} \bigg) \bigg]
\end{eqnarray}
and
\begin{equation}
\langle \psi_j^\sigma |V| 00 \rangle = {i g x_{00} \over \sqrt{2}} \, \delta_{j0}  -  \sigma {\Omega_0(0) \over 2 \sqrt{2}} \, \delta_{j1}.
\end{equation}

\subsection{Dressed state propagator}

In quantum computing applications one will often be interested in calculating transition amplitudes of the form
\begin{equation}
\langle {\rm f} | e^{-i H t/\hbar} | {\rm i} \rangle,
\end{equation}
where $| {\rm i} \rangle$ and $| {\rm f} \rangle$ are arbitrary initial and final states of the uncoupled qubit-resonator system. Expanding $| {\rm i} \rangle$ and $| {\rm f} \rangle$ in the dressed-state basis reduces the time-evolution problem to that of calculating the quantity
\begin{equation}
G_{jj'}^{\sigma \sigma'} \! (t) \equiv \langle \psi_{j}^{\sigma} |e^{-i H t/\hbar} |\psi_{j'}^{\sigma'}\rangle,
\label{propagator definition}
\end{equation}
as well as  $\langle \psi_{j}^{\sigma} |e^{-i H t/\hbar} | 00 \rangle$ and $\langle 00 |e^{-i H t/\hbar} |00 \rangle$.
$G_{jj'}^{\sigma \sigma'} \! (t)$ is a propagator in the dressed-state basis, and would be equal to $\delta_{\sigma \sigma'} \delta_{jj'} e^{-i W_j^\sigma t/\hbar}$ if $V$ were absent, that is, in the RWA.\cite{propagator footnote} Although it is possible to directly construct perturbative expressions for the propagator in the $|mn\rangle$ basis, the quantity defined in Eq.~(\ref{propagator definition}) turns out to be the simplest.

To be specific, we imagine preparing the system at $t=0$ in the state $|10\rangle$, which corresponds to the qubit in the excited state $m=1$ and the resonator in the ground state $n=0$. We then calculate the interaction-representation probability amplitude 
\begin{equation}
c_{mn}(t) \equiv e^{i E_{mn}t/\hbar} \langle mn| e^{-i H t/\hbar} |10\rangle
\end{equation}
for the system at a later time $t$ to be in the state $|mn\rangle$. Here $E_{mn} \equiv \epsilon_m + n \hbar \omega_0$. Inserting complete sets of the dressed states leads to
\begin{equation}
c_{00}(t) = \sum_{\sigma j} \langle \psi_j^\sigma |10\rangle \langle 00 | e^{-i H t/\hbar} |\psi_{j}^{\sigma}\rangle,
\end{equation}
and, for $mn \neq 00$,
\begin{widetext}
\begin{equation}
c_{mn}(t) = e^{i E_{mn} t/\hbar} \sum_{j=0}^\infty 
\begin{pmatrix} \langle \psi_j^+ |mn \rangle \cr  \langle \psi_j^- |mn \rangle \end{pmatrix}^\dagger
 \begin{pmatrix} G_{j0}^{++} & G_{j0}^{+-} \cr G_{j0}^{-+} & G_{j0}^{--} \end{pmatrix} \begin{pmatrix} \langle \psi_0^+ |10 \rangle \cr \langle \psi_0^- |10 \rangle \end{pmatrix}.
\label{cmn equation}
\end{equation}

Using the relations
\begin{equation}
|0n\rangle = {1 \over \sqrt{2}} \big[ |\psi_{n-1}^+\rangle + |\psi_{n-1}^-\rangle \big] \ \ \ \ ({\rm for} \ n \neq 0)
\end{equation}
and
\begin{equation}
|1n\rangle = {i \over \sqrt{2}} \big[ |\psi_n^+\rangle - |\psi_n^-\rangle \big] ,
\end{equation}
we obtain
\begin{equation}
c_{01}(t) = {\textstyle{i \over 2}} e^{i E_{01} t/\hbar} 
\begin{pmatrix} 1 \cr 1 \end{pmatrix}^\dagger \begin{pmatrix} G_{00}^{++} & G_{00}^{+-} \cr G_{00}^{-+} & G_{00}^{--} \end{pmatrix} \begin{pmatrix} 1 \cr -1 \end{pmatrix}
\label{c01 equation}
\end{equation}
and
\begin{equation}
c_{10}(t) = {\textstyle{1 \over 2}} e^{i  E_{10} t/\hbar} 
\begin{pmatrix} 1 \cr -1 \end{pmatrix}^\dagger \begin{pmatrix} G_{00}^{++} & G_{00}^{+-} \cr G_{00}^{-+} & G_{00}^{--} \end{pmatrix} \begin{pmatrix} 1 \cr -1 \end{pmatrix}.
\label{c10 equation}
\end{equation}
So far everything is exact within the model defined in Eq.~(\ref{full Hamiltonian}).

To proceed, we expand the dressed-state propagator in a basis of {\it exact} eigenstates $|\Psi_\alpha \rangle$ of $H$, leading to
\begin{equation}
G_{jj'}^{\sigma \sigma'} \! (t) = \sum_\alpha  \langle \psi_j^\sigma | \Psi_\alpha \rangle \, \langle \psi_{j'}^{\sigma'} | \Psi_\alpha \rangle^* \, e^{-i {\cal E}_\alpha t/\hbar}.
\label{propagator eigenfunction expansion}
\end{equation}
Here ${\cal E}_\alpha$ is the energy of stationary state $|\Psi_\alpha\rangle$. The propagator is an infinite sum of periodic functions of time. We approximate this quantity by evaluating the $|\Psi_\alpha\rangle$ and  ${\cal E}_\alpha$ perturbatively in the dressed-state basis. 

The leading-order corrections to the dressed-state energies are of order $V^2$. We obtain
\begin{eqnarray}
{\cal E}_{00} &=& - \sum_\sigma \bigg[ {|\langle \psi_0^\sigma |V| 00\rangle |^2 \over W_0^\sigma} + { |\langle \psi_1^\sigma |V| 00\rangle |^2 \over W_1^\sigma} \bigg] \nonumber \\
&=& -{g^2 \over 2} \sum_\sigma \bigg( {x_{00}^2 \over W_0^\sigma} + {x_{01}^2 \over W_1^\sigma} \bigg),
\end{eqnarray}
and
\begin{equation}
{\cal E}_{j \sigma} = W_j^\sigma + {|\langle \psi_j^\sigma |V| 00\rangle |^2 \over W_j^\sigma} + \sum_{j' \neq j,\sigma'} {|\langle \psi_j^\sigma |V|  \psi_{j'}^{\sigma'} \rangle |^2 \over W_j^\sigma - W_{j'}^{\sigma'}} .
\end{equation}
We will also need the {\it second}-order eigenfunctions, which, for a perturbation having no diagonal dressed-state matrix elements, are
\begin{equation}
|\Psi_{00}\rangle = A_{00} \bigg[  |00\rangle - \sum_{j \sigma} {\langle \psi_j^\sigma |V|00\rangle \over W_j^\sigma} \, |\psi_j^\sigma \rangle
+  \sum_{j j' \sigma \sigma'} { \langle \psi_j^\sigma |V| \psi_{j'}^{\sigma'}\rangle \langle \psi_{j'}^{\sigma'} |V| 00\rangle \over W_j^\sigma  W_{j'}^{\sigma'}} \, |\psi_j^\sigma \rangle  \bigg]
\end{equation}
and
\begin{eqnarray}
|\Psi_{\! j \sigma} \rangle = A_{j \sigma} \bigg[ |\psi_j^\sigma \rangle &+& {\langle 00 |V|\psi_j^\sigma \rangle \over W_j^\sigma} \, |00 \rangle 
+  \sum_{j' \sigma' \neq j\sigma} \bigg( {\langle \psi_{j'}^{\sigma'} |V|\psi_j^\sigma \rangle \over W_j^\sigma -W_{j'}^{\sigma'}} \, |\psi_{j'}^{\sigma'}\rangle
+  { \langle 00 |V| \psi_{j'}^{\sigma'}\rangle \langle \psi_{j'}^{\sigma'} |V| \psi_j^\sigma \rangle \over W_j^\sigma  (W_j^\sigma - W_{j'}^{\sigma'}) } \, |00 \rangle
+  { \langle \psi_{j'}^{\sigma'} |V| 00 \rangle \langle 00 |V| \psi_j^\sigma \rangle \over W_j^\sigma  (W_j^\sigma - W_{j'}^{\sigma'}) } \, |\psi_{j'}^{\sigma'} \rangle \bigg) \nonumber \\
&+&   \sum_{j' \sigma' \neq j\sigma} \sum_{j'' \sigma'' \neq j\sigma} { \langle \psi_{j'}^{\sigma'} |V| \psi_{j''}^{\sigma''}\rangle \langle \psi_{j''}^{\sigma''} |V|  \psi_j^\sigma \rangle \over (W_j^\sigma - W_{j'}^{\sigma'})(W_j^\sigma - W_{j''}^{\sigma''})  } \, |\psi_{j'}^{\sigma'} \rangle  \bigg],
\end{eqnarray}
where $A_{00}$ and the $A_{j \sigma}$ are normalization factors.

Writing out Eq.~(\ref{propagator eigenfunction expansion}) explicitly as
\begin{equation}
G_{j j'}^{\sigma \sigma'} \! (t) = \langle \psi_j^\sigma | \Psi_{00} \rangle  \langle \psi_{j'}^{\sigma'} | \Psi_{00} \rangle^* \, e^{-i {\cal E}_{00} t/\hbar} 
+ \sum_{\bar{\jmath} \bar{\sigma}}  \langle \psi_j^\sigma | \Psi_{\bar{\jmath} \bar{\sigma}} \rangle  \langle \psi_{j'}^{\sigma'} | \Psi_{\bar{\jmath} \bar{\sigma}} \rangle^* \, e^{-i {\cal E}_{\bar{\jmath} \bar{\sigma}} t/\hbar},
\end{equation}
and again making use of the fact that the matrix elements of $V$ diagonal in $j$ vanish, leads to
\begin{eqnarray}
G_{00}^{\sigma \sigma'} \! (t) &=&  \delta_{\sigma \sigma'} A_{0\sigma}^2 e^{-i {\cal E}_{0\sigma} t/\hbar}  
+ A_{00}^2 { \langle \psi_{0}^{\sigma} |V|00 \rangle \langle \psi_{0}^{\sigma'} |V|00 \rangle^* \over W_0^{\sigma} \, W_0^{\sigma'} } e^{-i {\cal E}_{00} t/\hbar} \nonumber \\
&+& A_{0\sigma}^2 (1-\delta_{\sigma \sigma'}) \bigg[ { \langle \psi_{0}^{\sigma'} |V|00 \rangle^* \langle 00 |V|\psi_{0}^{\sigma} \rangle^* \over W_0^{\sigma} (W_0^{\sigma}- W_0^{\sigma'}) } 
+ \sum_{\bar{\jmath} \neq 0,\bar{\sigma}} { \langle \psi_{0}^{\sigma'} |V| \psi_{\bar{\jmath}}^{\bar{\sigma}} \rangle^* \langle \psi_{\bar{\jmath}}^{\bar{\sigma}} |V|\psi_{0}^{\sigma} \rangle^* \over (W_0^{\sigma} - W_0^{\sigma'}) (W_0^{\sigma}- W_{\bar{\jmath}}^{\bar{\sigma}}) } \bigg] e^{-i {\cal E}_{0\sigma}t/\hbar} \nonumber \\
&+& A_{0\sigma'}^2 (1-\delta_{\sigma \sigma'}) \bigg[ { \langle \psi_{0}^{\sigma} |V|00 \rangle \langle 00 |V|\psi_{0}^{\sigma'} \rangle \over W_0^{\sigma'} (W_0^{\sigma'} - W_0^{\sigma}) } 
+ \sum_{\bar{\jmath} \neq 0,\bar{\sigma}} { \langle \psi_{0}^{\sigma} |V| \psi_{\bar{\jmath}}^{\bar{\sigma}} \rangle \langle \psi_{\bar{\jmath}}^{\bar{\sigma}} |V|\psi_{0}^{\sigma'} \rangle \over (W_0^{\sigma'} - W_0^{\sigma}) (W_0^{\sigma'} - W_{\bar{\jmath}}^{\bar{\sigma}}) } \bigg] e^{-i {\cal E}_{0\sigma'}t/\hbar} \nonumber \\
&+& \sum_{\bar{\jmath} \neq 0,\bar{\sigma}} A_{\bar{\jmath} \bar{\sigma}}^2 { \langle \psi_0^\sigma |V| \psi_{\bar{\jmath}}^{\bar{\sigma}} \rangle \langle \psi_0^{\sigma'} |V| \psi_{\bar{\jmath}}^{\bar{\sigma}} \rangle^*  \over (W_{\bar{\jmath}}^{\bar{\sigma}} - W_0^{\sigma})(W_{\bar{\jmath}}^{\bar{\sigma}} - W_0^{\sigma'})   }  e^{-i {\cal E}_{\bar{\jmath} \bar{\sigma}} t/\hbar} + {\cal O}(V^3),
\end{eqnarray}
or
\begin{eqnarray}
G_{00}^{\sigma \sigma'} \! (t) &=& \delta_{\sigma \sigma'} A_{0\sigma}^2 e^{-i {\cal E}_{0\sigma} t/\hbar}  
+ A_{00}^2 {g^2 x_{00}^2 \over 2 \,W_0^{\sigma} \, W_0^{\sigma'} } e^{-i {\cal E}_{00} t/\hbar}
+ A_{0\sigma}^2 \, {1-\delta_{\sigma \sigma'} \over W_0^\sigma - W_0^{\sigma'}} \bigg[ { g^2 x_{00}^2 \over 2 W_0^\sigma}  
+ \sum_{\bar{\sigma}} \bigg({ g^2 X_{\sigma \bar{\sigma}} X_{\sigma' \bar{\sigma}}   \over 4(W_0^\sigma - W_{1}^{\bar{\sigma}})  } + { g^2 x_{01}^2 \over 2  (W_0^\sigma - W_2^{\bar{\sigma}}) } \bigg) \bigg]   \nonumber \\
&\times& e^{-i {\cal E}_{0\sigma} t/\hbar} 
- A_{0\sigma'}^2 \, {1-\delta_{\sigma \sigma'} \over W_0^\sigma - W_0^{\sigma'}} \bigg[ { g^2 x_{00}^2 \over 2 W_0^{\sigma'} } + \sum_{\bar{\sigma}} 
\bigg({ g^2 X_{\sigma \bar{\sigma}} X_{\sigma' \bar{\sigma}}  \over 4(W_0^{\sigma'} - W_{1}^{\bar{\sigma}})  } 
+ { g^2 x_{01}^2 \over 2  (W_0^{\sigma'} - W_2^{\bar{\sigma}}) } \bigg) \bigg]  e^{-i {\cal E}_{0\sigma'} t/\hbar} \nonumber \\
&+& {1 \over 4} \sum_{\bar{\sigma}} A_{1 \bar{\sigma}}^2 { g^2 X_{\sigma \bar{\sigma}} X_{\sigma' \bar{\sigma}}  \over (W_{1}^{\bar{\sigma}} - W_0^{\sigma})(W_{1}^{\bar{\sigma}} - W_0^{\sigma'})  } \,  e^{-i {\cal E}_{1 \bar{\sigma}} t/\hbar} 
+ {1 \over 2} \sum_{\bar{\sigma}} A_{2 \bar{\sigma}}^2 { g^2 x_{01}^2 \over  (W_{2}^{\bar{\sigma}} - W_0^{\sigma}) (W_{2}^{\bar{\sigma}} - W_0^{\sigma'}) }  \, e^{-i {\cal E}_{2 \bar{\sigma}} t/\hbar},
\end{eqnarray}
where
\begin{equation}
X_{\sigma \sigma'} \equiv \sqrt{2} \, x_{00} + \sigma \sigma^\prime \, x_{11}.
\end{equation}
Note that there are no order $V$ corrections to the dressed-state propagator. Because of this property, the leading order corrections are of order $V^2$, and it is therefore necessary to use second-order perturbative eigenfunctions to obtain all such second-order terms.

Finally, we note that the normalization constants are simply
\begin{eqnarray}
A_{00} = \bigg[1+ {g^2 \over 2} \sum_\sigma \bigg( {x_{00}^2 \over (W_0^\sigma)^2} +  {x_{01}^2 \over (W_1^\sigma)^2} \bigg) \bigg]^{-{1\over 2}},
 \end{eqnarray}
\begin{eqnarray}
A_{0\sigma} = \bigg[1+ {g^2 x_{00}^2 \over 2(W_0^\sigma)^2} + {g^2 \over 4} \sum_{\sigma'} \bigg( {(\sqrt{2} \, x_{00} + \sigma \sigma' \, x_{11})^2 \over (W_0^\sigma-W_1^{\sigma'})^2} +  {2 \, x_{01}^2 \over (W_0^\sigma - W_2^{\sigma'})^2} \bigg) \bigg]^{-{1\over 2}},
 \end{eqnarray}
\begin{eqnarray}
A_{1\sigma} = \bigg[1+ {g^2 x_{01}^2 \over 2(W_1^\sigma)^2} + {g^2 \over 4} \sum_{\sigma'} \bigg( {(\sqrt{2} \, x_{00} + \sigma \sigma' \, x_{11})^2 \over (W_1^\sigma - W_0^{\sigma'})^2} + {(\sqrt{3} \, x_{00} + \sqrt{2} \, \sigma \sigma' \, x_{11})^2 \over (W_1^\sigma - W_2^{\sigma'})^2} + {3 \, x_{01}^2 \over (W_1^\sigma - W_3^{\sigma'})^2} \bigg) \bigg]^{-{1\over 2}},
 \end{eqnarray}
and
\begin{eqnarray}
A_{2\sigma} = \bigg[1+ {g^2 \over 4} \sum_{\sigma'} \bigg( {2 \, x_{01}^2 \over (W_2^\sigma - W_0^{\sigma'})^2} + {(\sqrt{3} \, x_{00} + \sqrt{2} \, \sigma \sigma' \, x_{11})^2 \over (W_2^\sigma - W_1^{\sigma'})^2} + {( 2 \, x_{00} + \sqrt{3} \, \sigma \sigma' \, x_{11})^2 \over (W_2^\sigma - W_3^{\sigma'})^2} + {4 \, x_{01}^2 \over (W_2^\sigma - W_4^{\sigma'})^2} \bigg) \bigg]^{-{1\over 2}}.
 \end{eqnarray}

\end{widetext}

\section{TOWARDS INFORMATION PROCESSING WITH STRONG COUPLING}

In this section, we test our perturbed dressed-state method for the case of a finite-dimensional single-qubit, five-phonon system. The junction has
parameters $E_{\rm J} = 43.1 \, {\rm meV}$ and $E_{\rm c} = 53.4 \, {\rm neV},$ corresponding to that of Ref.~\onlinecite{Martinis 2002}. The resonator has a frequency $\omega_0/2 \pi$ of $10 \, {\rm GHz},$ and the interaction strength $g$ varies from weak $(g \ll \Delta \epsilon)$ to strong $(g \approx \Delta \epsilon)$ coupling. The bias current is chosen to make the the system exactly in resonance, and this bias is sufficiently smaller than the critical current so that the junction states are well approximated by harmonic oscillator eigenfunctions. The Hamiltonian for this system is diagonalized numerically, and the probability amplitudes $c_{mn}(t)$ are calculated exactly, providing both a test of the accuracy of the analytic perturbative solutions and an estimate of the range of interaction strengths $g$ for which it is valid. Setting the initial state to be $c_{mn}(0) = \delta_{m1} \delta_{n0}$, as assumed previously, we simulate the transfer of a qubit from the Josephson junction to the resonator, by leaving the systems in resonance for half a vacuum Rabi period $\pi \hbar / g |x_{01}|.$\cite{Cleland and Geller 2004,Geller and Cleland}

\begin{figure}
\includegraphics[width=8.0cm]{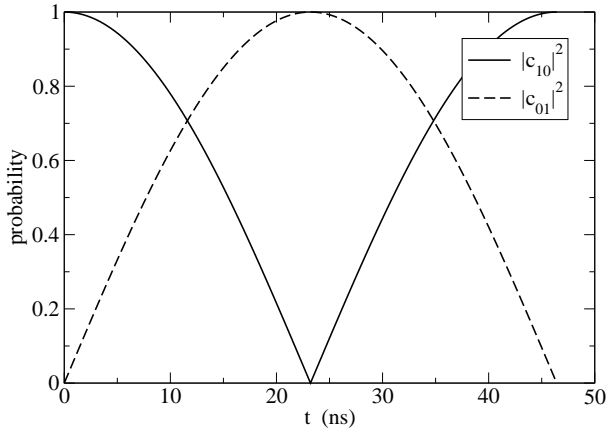}
\caption{\label{g=0.03 figure}Time evolution of probabilities $|c_{10}|^2$ and $|c_{01}|^2$ for the weakly coupled case of $g/\Delta \epsilon = 0.03$. Here the exact, RWA, and dressed-state perturbative results are essentially equivalent.}
\end{figure}

\begin{figure}
\vskip 0.1in
\includegraphics[width=8.0cm]{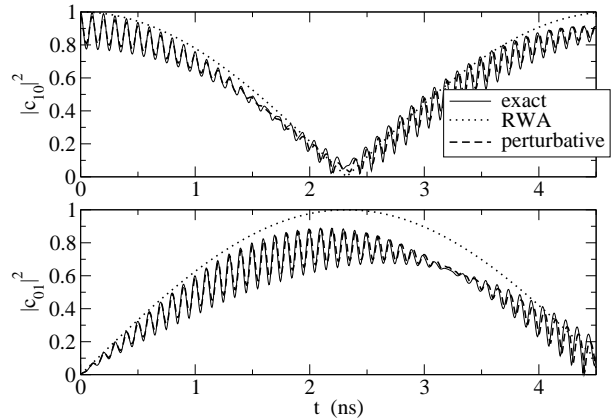}
\caption{\label{g=0.30 figure}Probabilities $|c_{10}|^2$ and $|c_{01}|^2$ for the strongly coupled case $g/\Delta \epsilon = 0.30$. Here there are large deviations from the RWA behavior, which are correctly accounted for by the dressed-state perturbative method. Note the ten-fold increase in transfer speed compared with that of Fig.~\ref{g=0.03 figure}.}
\end{figure}

\begin{figure}
\vskip 0.2in
\includegraphics[width=8.0cm]{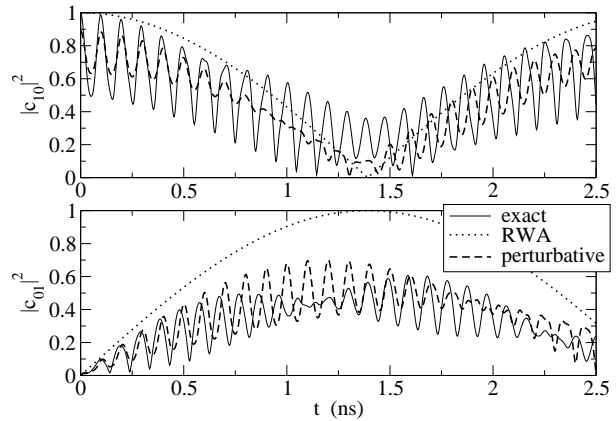}
\caption{\label{g=0.50 figure}Probabilities $|c_{10}|^2$ and $|c_{01}|^2$ for the strongly coupled case $g/\Delta \epsilon = 0.30$. Here both the RWA and dressed-state perturbative approximations fail.}
\end{figure}

Figures~\ref{g=0.03 figure}, \ref{g=0.30 figure}, and \ref{g=0.50 figure}, show the time evolution of the occupation probabilities $|c_{10}|^2$ and $|c_{01}|^2 \! ,$ for different values of $g$. In Fig.~\ref{g=0.03 figure}, we plot the results for very weak coupling, $g/\Delta \epsilon = 0.03$. The evolution takes the junction qubit $|1\rangle_{\rm JJ}$ and transfers it to and from the resonator periodically. The exact, RWA, and dressed-state perturbative results are all the same to within the thickness of the lines shown in Fig.~\ref{g=0.03 figure}. Thus, for this value of $g$, the RWA is extremely accurate.

In Fig.~\ref{g=0.30 figure}, we plot the probabilities for stronger coupling, $g/\Delta \epsilon = 0.30$. For this coupling strength, the RWA is observed to fail. For example, the RWA predicts a perfect state transfer between the junction and the resonator, and does not exhibit the oscillations present in the exact solution. The dressed-state perturbative approximation does correctly capture these oscillations. In Fig.~\ref{g=0.50 figure}, we show the same quantities for the case $g/\Delta \epsilon = 0.5$. At this coupling strength, both the RWA and the dressed-state perturbative approximation break down.

\section{STATE TRANSFER FIDELITY}

In this final section, we briefly investigate to what extent we may increase the junction-resonator coupling $g$, and still have an accurate state transfer from the Josephson junction to the resonator. As before, we start at time $t \! = \! 0$ in the state $|10\rangle$. In order to define the fidelity of the state transfer operation, we first determine the time $t_{\rm min}$ of the minimum of the probability $|c_{10}(t)|^2 \! .$ Recall that $|c_{10}|^2$ is the probability that the junction is in the $m \! = \! 1$ excited qubit state and the resonator is in the $n \! = \! 0$ vacuum state. 

\begin{figure}
\includegraphics[width=8.0cm]{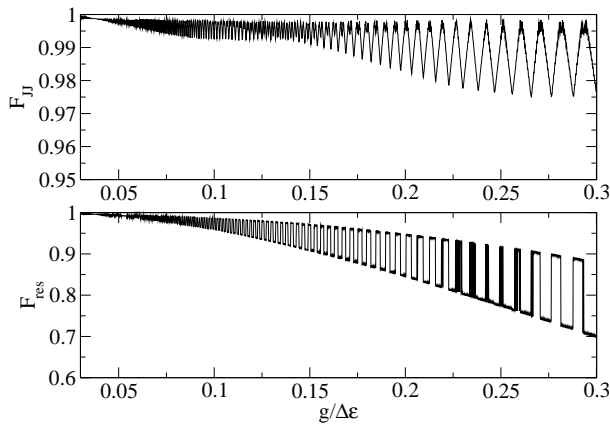}
\caption{\label{fidelity figure} The fidelity of a state transfer from Josephson junction to nanomechanical resonator as a function of interaction strength $g$. Note the difference in scale between the upper and lower curves.}
\end{figure}

It will be convenient to define two fidelities: $F_{\rm JJ} \equiv 1-|c_{10}(t_{\rm min})|^2$ is the fidelity (or, more precisely, the fidelity squared) for the junction, and $F_{\rm res} \equiv |c_{01}(t_{\rm min})|^2$ is the squared fidelity for the resonator.\cite{fidelity footnote} These quantities are different because of leakage to other states; however, in the RWA limit, they are both equal to unity. $F_{\rm JJ}$ measure the success of de-exciting the qubit, and $F_{\rm res}$ measures the success of exciting the resonator. In Fig.~\ref{fidelity figure} we plot $F_{\rm JJ}$ and $F_{\rm res}$ as a function of $g$. Typically, the junction fidelity $F_{\rm JJ}$ remains close to unity, with some oscillations, for all couplings. This behavior is a consequence of the fact that there is always a time where $|c_{10}|^2$ becomes small, as is evident in Figs.~\ref{g=0.30 figure} and \ref{g=0.50 figure}. However, because of leakage to other states, the resonator fidelity $F_{\rm res}$ decreases significantly (again with oscillations due to the``switching'' of $t_{\rm min}$ with $g$) with increasing interaction strength.
The lower curve in  Fig.~\ref{fidelity figure} shows that $F_{\rm res} \ge  90\%$ is possible with $g = 0.15 \, \Delta \epsilon$, which allows a state transfer in under $5 \, {\rm ns}$.

\section{DISCUSSION}

We have developed a theoretical technique to analytically calculate the leading-order perturbative corrections to the RWA or Jaynes-Cummings Hamiltonian for a quantum two-level system linearly coupled to a harmonic oscillator or single-mode boson field, a model central to many current quantum computing architectures. Such corrections are necessary to treat the fast information-processing regime where the interaction strength approaches the qubit level spacing. The method was applied to a current-biased Josephson junction coupled to a piezoelectric nanoelectromechanical disk resonator, and good agreement with exact numerical results was obtained.

\acknowledgments

It is a pleasure to thank Emily Pritchett and Steve Lewis for useful discussions. ANC was supported by the DARPA/DMEA Center for Nanoscience Innovation for Defence. MRG was supported by the National Science Foundation under CAREER Grant No.~DMR-0093217, and by the Research Corporation.


\begin{thebibliography}{99}
\bibitem{generalization footnote} If the higher lying junction states $m=2,3,\dots$ are not initially occupied, and the resonator also starts in the ground state, only the qubit states $m=0,1$ are coupled by ``energy conserving" terms. Therefore, the higher lying junction states $m=2,3,\dots$ can be treated with ordinary perturbation theory in $g$.
\bibitem{propagator footnote} An alternative representation for the propagator is
$$
G_{jj'}^{\sigma \sigma'} \! (t) = e^{-i W_j^\sigma t/\hbar} \, \langle \psi_{j}^{\sigma} |T e^{-(i/\hbar) \int_0^t \! d\tau \, V(\tau)} |\psi_{j'}^{\sigma'}\rangle, 
$$
where $V(t) \! \equiv \! e^{i H_{\rm JC} t/\hbar} V e^{-i H_{\rm JC} t/\hbar}$, and where $T$ is the time-ordering operator. However, directly computing $G_{jj'}^{\sigma \sigma'} \! (t)$ by a series or cumulant expansion in $V$ leads to a result that is valid only for short times.
\bibitem{fidelity footnote} Because the desired final state is $|01\rangle$, overall squared fidelity of the state transfer operation is simply $F_{\rm res}$.
\end{thebibliography}
\end{document}